\documentclass[aps,showpacs,nofootinbib]{revtex4}
\usepackage{epsf,latexsym}

\begin{document}

\title{A note on the connection between the universal relaxation bound and
the covariant entropy bound}
\author{Alessandro Pesci}
\email{pesci@bo.infn.it}
\affiliation
{INFN, Sezione di Bologna, Via Irnerio 46, I-40126 Bologna, Italy}

\begin{abstract}
A recently formulated 
universal lower-bound to the characteristic relaxation times
of perturbed thermodynamic systems,
derived from quantum information theory
and (classical) thermodynamics and
known to be saturated 
for (certain) black holes,
is investigated 
in the light of the
gravity/thermodynamics connection.
A statistical-mechanical property, 
unrelated to gravity, 
essential for the validity
of the generalized covariant entropy bound,
namely the existence of a lower-limiting
value $l^*$ for the size of thermodynamic
systems, 
is found to provide a way to understand
this universal relaxation bound,
thus regardless of the kind of foundations
(i.e. whether conventional or information-based)
of the statistical-mechanical description.
As a by-product
an example of a conventional system
(i.e. not a black hole) 
seemingly saturating the universal relaxation
bound is provided.

\end{abstract}

\pacs{04.20.Cv, 04.70.Dy, 04.62.+v, 05.20.-y, 05.30.-d, 05.70.-a}

\maketitle

$ $
Recently, a precise formulation 
of a universal lower-bound to the relaxation times
of perturbed thermodynamic systems 
has been proposed \cite{Hod}. 
For a given thermodynamic system,
if $\tau$ is the time
characterising the rate of change of a quantity
which starts from some non-equilibrium value,
the expression of the bound,
named TTT (Time times Temperature) by its discoverer, 
is

\begin{eqnarray}\label{HodBound}
\tau \geq \frac{\hbar}{\pi k T} \equiv \tau_{min},
\end{eqnarray}
being $T$ the temperature of the system
(and $k$ and $\hbar$ the Boltzmann and (reduced) Planck constants).
Its derivation hinges on quantum information theory
and (classical) thermodynamic considerations \cite{Hod}.
This same bound however, 
at least for certain systems,
appears to be reachable through
a different path, not directly related to quantum information theory,
and the aim of this note is to briefly
describe this alternative approach.
The derivation of bound (\ref{HodBound}) 
does not involve gravity (see \cite{Hod} 
and \cite{BekensteinInfo, Bremermann}).
At the same time however
the only known systems
able to attain the bound
are (certain) black holes \cite{Hod, Gruzinov, Hod2}.  
A by-product of the approach followed here will be to show
that also some conventional thermodynamic systems
are seemingly able to saturate the bound.

Once again black holes signal a connection between
gravity and thermodynamics,
the main evidence for it
coming from the laws of black hole mechanics \cite{Bardeen},
the assignement of an entropy to black holes
\cite{Bekenstein, Hawking}
and the formulation of a 
generalized second law \cite{BekensteinGSL, Hawking};
one may wonder if among these 
thermodinamic/gravitational laws a clue
can be hidden able to bring to bound (\ref{HodBound}).
Since in bound (\ref{HodBound}) gravity has no role,
this clue should have a peculiar nature:
to be required for the thermodynamical laws of gravity 
to hold
and to possess an intrinsic meaning not depending on gravity.

The generalized second law
has been shown to be a consequence
of the generalized Bousso bound
--a generalization of the Bousso covariant
entropy bound \cite{Bousso}--
provided the ordinary second law is assumed to hold
\cite{FlanaganMarolfWald}.
This generalized bound
asserts that the entropy $S$ on a 
lightsheet 
(a null hypersurface, 
generated by non-expanding light rays
emanating orthogonally from an assigned spacelike
2-surface)
associated with a given
2-suface with area $A$
and truncated on another spacelike 2-surface
with area $A^\prime (\leq A)$ satisfies
(in Planck units, the units we will use from now on
unless explicitly stated otherwise)

\begin{eqnarray}\label{GenBBound}
S \leq \frac{1}{4} (A-A^\prime).
\end{eqnarray}
In addition of having the generalized second law included in it,
this entropy bound subsumes also the various proposed entropy bounds,
such as the universal entropy bound of Bekenstein \cite{BekensteinBound}
or the holographic bound 
\cite{tHooft, Susskind} 
(closely related to the so-called
holographic principle \cite{tHooft, Susskind}, 
roughly the idea that the physics inside a region
of spacetime can be fully described by degrees of freedom
living on the boudary),
while avoiding their limitations.
It seems thus to stay at the hearth
of the connection between gravity and 
thermodynamics/statistical physics,
somehow summarizing its content
(besides the thermodynamic imprint
directly grasped
in Einstein's field equations
\cite{Jacobson} and in the action
\cite{Padmanabhan}).
If a clue bringing to (\ref{HodBound}) 
somewhere in thermo-gravitational laws does exist at all,
it should not lie outside
this entropy bound. 

In \cite{PesciProof} a novel proof of
the bound (\ref{GenBBound}) has been given
(other proofs are in \cite{FlanaganMarolfWald} and 
\cite{BoussoFlanaganMarolf, Strominger})
in which,
reconducting
the validity of the bound
as applied to a generic thermodynamic system
to its validity on each thin slice
which the system can be thought to be made of,
at least for ultra-relativistic ideal fluids
with vanishing chemical potential $\mu$
a condition is set, 
in terms of local variables
and the thickness of the slice,
sufficient and necessary for the validity of the bound.
This condition,
generalized in \cite{PesciLength} 
to address the case of generic ideal fluids,
reads

\begin{eqnarray}\label{microBousso}
s \leq \pi l (\rho + p),
\end{eqnarray}
where $s$, $\rho$ and $p$
are local entropy density, energy density and pressure respectively
and $l$ is the thickness of the slice. 
This formula looks similar
to some of the conditions found to be sufficient to prove
the bound (\ref{GenBBound}) in other derivations
of it \cite{FlanaganMarolfWald, BoussoFlanaganMarolf}. 
At the end all these conditions correspond to some reformulation
of the Bekenstein bound.

The result (\ref{microBousso}) can be reinterpreted
\cite{PesciProof, PesciLength}
as showing that the generalized bound (\ref{GenBBound})
is universally satisfied {\it iff}
a minimal length scale $l^*$ exists
for the size $l$ of the systems for which a meaningful
notion of statistical entropy can be given,
so that always

\begin{eqnarray}\label{l_star}
l \geq l^* \equiv \frac{1}{\pi} \ \frac{s}{\rho + p}
= \frac{1}{\pi T} \ \left( 1 - \frac{\mu n}{\rho + p} \right),
\end{eqnarray}  
where $n$ is local number density
and in $\mu$ the (possible) rest-energy 
of constituent particles is included.
The expression for $l^*$ turns out to be very similar
to the expression (\ref{HodBound}) above for $\tau_{min}$,
in particular when $\mu = 0$.
\footnote{The noticing of this point is due to
S. Hod (see also \cite{PesciLength}).}
In this case we have

\begin{eqnarray}\label{l_photons}
l^* = \frac{1}{\pi T} = \frac{\hbar c}{\pi k T},
\end{eqnarray}
where in the last equality 
all the constants have been 
inserted ($c$ is the speed of light),
so that the expressions for $l^*$ and $\tau_{min}$
are absolutely similar
(actually coincident, in Planck units).
Analogously to Hod's TTT bound,
the bound (\ref{l_star}), as well as
the condition (\ref{microBousso}), has no reference
to gravity
(yet $l^*$ will be in general much larger than
the Planck length \cite{PesciLength}). 
As discussed in \cite{PesciLength}
the origin of $l^*$ appears to be entirely
statistical-mechanical
(and this is exactly what happens also for
the Bekenstein bound \cite{BekensteinBound});
the emerging scenario is that
quantum mechanics, requiring the existence
of a minimal size $l_{min} \geq l^*$ for thermodynamic systems
(usually by far satisfied),
ultimately determined by the spatial quantum uncertainty
of the constituent particles
(or the maximum between this and their size
in case of composite objects),
protects the generalized Bousso bound (\ref{GenBBound}).
Conversely, if
the bound (\ref{GenBBound})
is axiomatically assumed, 
quantum mechanics is required to emerge, 
in order to fix the minimal length scale 
accompanying necessarily the bound.
\footnote{Reasoning on single-particle systems,
in \cite{BoussoFlat} the role of quantum
mechanics as protecting the Bekenstein bound 
or being predicted by it, 
has been stressed.}

The existence of lower-limiting sizes 
for the statistical-mechanical systems
is just what
provides a connection between the entropy
bound (\ref{GenBBound}) and the  
relaxation bound (\ref{HodBound}),
thus the clue bringing to (\ref{HodBound}).
Assuming in fact that 
the time $\tau$ characterising
the rate of change of an assigned quantity
(when starting from some non-equilibrium value)
is given by the time
it takes a perturbation to propagate
through the system 
(see for instance \cite{Landau}),
for slices of thickness $l_{min}$ 
we should expect 
$\tau = l_{min}/c_s$,
where $c_s$ is sound velocity,
and this becomes

\begin{eqnarray}\label{tau}
\tau = l_{min}
\end{eqnarray}
when perturbations propagate at the speed of light.
Two elements cause
$\tau$ to be not lower than $l^*$ for the assigned thermodynamic
conditions, namely the fact that, accepting the scenario above,
$l_{min} \geq l^*$ ($l_{min} \gg l^*$ usually)
and also that $c_s \leq 1$.
This suggests that even if in general
$l^*$ can be smaller than $1/\pi T$
(when $\mu > 0$ actually),
Hod's TTT bound can be protected
by the pertinent
$l_{min}$ and $c_s$ values. 

What we find quite interesting and worth stressing here
is what happens for
a photon gas
or, more generally, for a gas of ultra-relativistic
particles with $\mu = 0$.
If in fact, as discussed in \cite{PesciLength}, 
photon-gas slices can be chosen
with limiting thickness 
$l = l_{min} = l^* = \frac{1}{\pi T}$,
they will saturate
the generalized Bousso bound \cite{PesciLength},
but from (\ref{tau}) they do saturate
also Hod's TTT bound,
adding to the known example of
black holes \cite{Hod, Gruzinov, Hod2}.

The discussion in \cite{PesciLength} on photon gases
shows moreover that it is not possible
to choose slices with thickness smaller than $l^*$
without destroying the thermodynamical
conditions in the slice.
This could then be considered as an argument
bringing to Hod's TTT bound 
(actually for photon gases or
for ultra-relativistic systems with $\mu=0$)
totally internal to statistical mechanics,
without the need of explicit reference
to quantum information theory.
At the end the fact that a thermodynamic result can be understood
alternatively from quantum information theory
or from conventional statistical mechanics should 
not be surprising 
in the light of the strong evidence of
equivalence between
conventional and quantum-informational
based statistical mechanics (see for example \cite{Ben-Arieh}).

The generic existence of some universal lower limit 
for the characteristic time $\tau$
with inverse dependence on temperature 
is known to be required within 
conventional statistical mechanics \cite{Landau}
(as stressed in \cite{Ropotenko}).
The argument 
just given turns out simply 
to permit one
to focus on a precise expression
for this limit, 
\footnote{In \cite{Sachdev} 
the relaxation bound
in the form $\tau \geq C \frac{1}{T}$,
with $C$ a number of order unity, 
was first discussed and models have been considered
(describing quantum critical points),
for some of which $\tau$ approaches a few $1/T$.}
the TTT bound,
linking it to the generalized Bousso bound, exactly.
The key element for this is the existence of a lower-limiting
scale $l^*$ for the size of statistical systems,
required for the thermo-gravitational laws to hold.
This limiting scale appears to be understandable
within statistical mechanics alone
(regardless moreover of its foundations,
i.e. whether quantum-informational or conventional)
so that
its meaning, existence and value have no relation at all
to gravity.

I would like to thank Shahar Hod for the stimulating correspondence.


\begin{thebibliography}{00}

\bibitem{Hod} S. Hod,
Phys. Rev. D {\bf 75} (2007) 064013,
gr-qc/0611004.

\bibitem{BekensteinInfo} J.D. Bekenstein,
Phys. Rev. Lett. {\bf 46} (1981) 623.

\bibitem{Bremermann} H.J. Bremermann,
in Proc. of Fifth Berkeley Symposium 
on Mathematical Statistics and Probability, 
edited by L. M. LeCam and J. Neyman 
(Univ. of California Press, Berkeley, 1967).

\bibitem{Gruzinov} A. Gruzinov,
arXiv:0705.1725.

\bibitem{Hod2} S. Hod,
Class. Quantum Grav. {\bf 24} (2007) 4235,
arXiv:0705.2306.

\bibitem{Bardeen} J.M. Bardeen, B. Carter and S.W. Hawking,
Commun. Math. Phys. {\bf 31} (1973) 161.

\bibitem{Bekenstein} J.D. Bekenstein,
Nuovo Cim. Lett. {\bf 4} (1972) 737;
Phys. Rev. D {\bf 7} (1973) 2333.

\bibitem{Hawking} S.W. Hawking,
Nature {\bf 248} (1974) 30;
Commun. Math. Phys. {\bf 43} (1975) 199.

\bibitem{BekensteinGSL} J.D. Bekenstein,
Phys. Rev. D {\bf 9} (1974) 3292.

\bibitem{FlanaganMarolfWald} $\acute {\rm E}$.$\acute {\rm E}$. 
Flanagan, D. Marolf and R.M. Wald,
Phys. Rev. D {\bf 62} (2000) 084035, hep-th/9908070. 

\bibitem{Bousso} R. Bousso,
JHEP07(1999) 004, hep-th/9905177;
JHEP06(1999) 028, hep-th/9906022;
Class. Quant. Grav. {\bf 17} (2000) 997, hep-th/9911002.

\bibitem{BekensteinBound} J.D. Bekenstein,
Phys. Rev. D {\bf 23} (1981) 287.

\bibitem{tHooft} G. 't Hooft,
in {\it Salamfest} (1993) 0284, 
gr-qc/9310026.

\bibitem{Susskind} L. Susskind,
J. Math. Phys. {\bf 36} (1995) 6377, 
hep-th/9409089.

\bibitem{Jacobson} T. Jacobson,
Phys. Rev. Lett. {\bf 75} (1995) 1260, 
gr-qc/9504004.

\bibitem{Padmanabhan} T. Padmanabhan,
Phys. Rept. {\bf 406} (2005) 49, gr-qc/0311036;
Int. J. Mod. Phys. D {\bf 15} (2006) 1659, gr-qc/0606061; 
AIP Conf. Proc. {\bf 939} (2007) 114, arXiv:0706.1654.

\bibitem{PesciProof} A. Pesci,
Class. Quantum Grav. {\bf 24} (2007) 6219, 
arXiv:0708.3729.

\bibitem{BoussoFlanaganMarolf} R. Bousso, 
$\acute {\rm E}$.$\acute {\rm E}$. Flanagan and D. Marolf,
Phys. Rev. D {\bf 68} (2003) 064001, 
hep-th/0305149.

\bibitem{Strominger} A. Strominger and D.M. Thompson,
Phys. Rev. D {\bf 70} (2004) 044007, 
hep-th/0303067.

\bibitem{PesciLength} A. Pesci,
Class. Quantum Grav. {\bf 25} (2008) 125005,
arXiv:0803.2642.

\bibitem{BoussoFlat} R. Bousso,
JHEP05(2004) 050,
hep-th/0402058.

\bibitem{Landau} L.D. Landau, E.M. Lifshitz and L.P. Pitaevskij,
{\it Statistical Physics} (Pergamon Press, Oxford, 1980).

\bibitem{Ben-Arieh} A. Ben-Naim,
{\it A farewell to entropy: statistical thermodynamics based 
on information} (World Scientific, Singapore, 2008).

\bibitem{Ropotenko} K. Ropotenko,
arXiv:0705.3625;
arXiv:0803.4489.

\bibitem{Sachdev} S. Sachdev,
{\it Quantum phase transitions}
(Cambridge University Press, New York, 1999).








\end{thebibliography}
\end{document}